# Pressure-induced Lifshitz and quantum phase transitions in electron-doped cuprate superconductor

Jinyu Zhao[1]*, Shu Cai[1]*, Zhaoyu Liu[2]*, Jianfeng Zhang[1]*, Shuaihang Sun[1], Pengyu Wang[2], Jing Guo[2], Yazhou Zhou[2], Shiliang Li[2]†, Fuyang Liu[1], Luhong Wang[3], Haozhe Liu[1]†, Yang Ding[1], Qi Wu[2], Tao Xiang[2]†, Richard L. Greene [4]† and Liling Sun[1, 2]†

[1]*Center for High Pressure Science & Technology Advanced Research, Beijing 100193, China*

[2]*Institute of Physics, Chinese Academy of Sciences, Beijing 100190, China*

[3]*Shanghai Key Laboratory of Material Frontiers Research in Extreme Environments, Shanghai Advanced Research in Physical Sciences, Shanghai 201203, China*

[4]*Maryland Quantum Materials Center, Department of Physics, University of Maryland, College Park, Maryland 20742, USA*

We report the first observations of a pressure-induced Lifshitz transition coupled with a quantum phase transition in electron-doped cuprate superconductor $Pr_{0.87}LaCe_{0.13}CuO_{4\pm\delta}$ (PLCCO), by combining high-pressure electrical resistance, Hall coefficient ($R_H$) and synchrotron X-ray diffraction (XRD) measurements at low temperatures. Our low-temperature Hall coefficient ($R_H$) measurements reveal that the $R_H$ decreases continuously and reaches zero at ~10 GPa (critical pressure of $P_{c1}$). Upon further compression beyond $P_{c1}$, $R_H$ unexpectedly changes its sign from negative to positive, signaling a reconstruction of the Fermi surface from electron-dominated to hole-dominated topology. Concurrently, the superconducting transition temperature ($T_c$) exhibits a monotonic suppression, vanishing completely at ~17.6 GPa (critical pressure of $P_{c2}$), where the system enters a non-superconducting metallic state. Our low temperature XRD measurements unequivocally demonstrate the absence of any structural phase transition across $P_{c1}$ and $P_{c2}$. Therefore, the sign change in $R_H$ at $P_{c1}$ is associated with a Lifshitz transition, which is never found in the compressed bulk electron- or hole-doped cuprate superconductors. Moreover, the quantum phase transition observed at $P_{c2}$ contrasts sharply with known high-pressure behavior of hole-doped cuprates, uncovering a fundamental difference on how pressure tunes the ground states of electron- versus hole-doped systems. These findings provide crucial insights into the different pressure responses on the interplay among Fermi surface topology, electronic correlations, and superconductivity between these two kinds of cuprate superconductors.

Although nearly four decades have passed since the discovery of high-$T_c$ superconductivity in copper-oxide (1), a unified understanding of its underlying mechanism has yet to be achieved (2-8). It is well known that the two control parameters, chemical doping and external pressure, can effectively tune the superconductivity. To date, doping effects on both hole- and electron-doped cuprates have been extensively investigated, providing valuable insights into the complex phenomena and physics of these materials (5, 8-19). Currently, what we know is that the undoped parent compounds of these high-$T_c$ superconductors are antiferromagnetic (AFM) Mott insulators. As electron or hole carriers are introduced through doping, the AFM state is suppressed and then the superconductivity emerges (1, 7, 8, 16, 18, 20, 21). By contrast, applying pressure to the undoped parent compound does not appear to induce a transition from the AFM to the superconducting state, as no such a case has been reported to date. Although pressure has a limited influence on the AFM state, its impact on the superconducting state of hole-doped cuprates is significant. Notable examples include the pressure-induced enhancement of $T_c$ (22-27) and the emergence of new phenomena in hole-doped cuprates (28-30). However, high-pressure studies on superconductivity of electron-doped cuprates remain remarkably unexplored (31-34). In this study, we report the first systematic investigation on the superconductivity of the electron-doped cuprate superconductor $Pr_{0.87}LaCe_{0.13}CuO_{4\pm\delta}$ (PLCCO), through high-pressure and low-temperature measurements of Hall coefficient, synchrotron X-ray diffraction, and resistance.

## Results

**Superconductivity Under Pressure.** First, we performed high-pressure resistance measurements on the superconducting $Pr_{0.87}LaCe_{0.13}CuO_{4\pm\delta}$ (PLCCO) single crystals. As shown in Fig. 1(a)-1(d) and Supplemental Information (SI), the superconducting transition temperature ($T_c$) of the five samples displays slight variation upon increasing pressure up to ~10 GPa, above which $T_c$ begins to decrease monotonically. Zero resistance is maintained up to 15.3 GPa. However, between 15.3 and 16.1 GPa, the samples lose zero-resistance (Fig.1(c) and 1(d)), indicating the suppression of the superconducting state. Upon further compression, the superconducting state is completely suppressed and replaced by a metallic state. This continuous evolution from a superconducting (SC) to a metallic (M) ground states provides strong evidence for a quantum phase transition (QPT).

Previous work by Butch et al. (35) established that such a quantum phase transition can be tuned by doping in the $La_{2-x}Ce_xCuO_4$ system, where the low-temperature normal-state resistivity evolves from a $T$-linear behavior in the superconducting doping regime to a $T^2$ dependence in the non-superconducting regime. However, our high-pressure results reveal a markedly different temperature dependence of the normal-state resistivity (see Fig. 1 and SI), suggesting that the effects of pressure and doping are not the same. This discrepancy warrants further investigation to elucidate the underlying physics.

**Hall Resistance Under High Pressure and Low Temperature.** Next, we conducted high-pressure Hall resistance ($R_{xy}$) measurements, a powerful technique for probing Fermi surface properties in a metal (36), by sweeping the magnetic field ($B$) from 0 to 5 T perpendicular to the *ab* plane of single-crystal sample 3 (S#3) and sample 4 (S#4) at 40 K. Unexpectedly, a sign change in $R_{xy}(B)$ from negative to positive was observed between 9.3-10.4 GPa for S#3 (Fig. 2(a)) and above 9.9 GPa for S#4 (Fig. 2(b)). Based on the results from the experiments conducted over the two runs, we estimate the critical pressure ($P_{c1}$) associated with this sign change to be around 10 GPa. It is found that, although electron carriers dominate conduction below 10 GPa, the contribution of hole carriers is significantly enhanced upon increasing pressure.

**High-Pressure and Low-Temperature Structure Information.** To investigate whether the pressure-induced change in the dominant carrier type and the suppression of superconductivity in PLCCO are associated with any pressure-induced structural phase transition, we perform the high-pressure and low-temperature synchrotron X-ray diffraction (XRD) measurements on the PLCCO sample (Fig. 3), which has not been performed before. The XRD patterns were collected at 40 K, the same as the temperature used for the Hall resistance measurements. In these experiments, the sample was cooled from 300 K down to 40 K, then the XRD patterns were collected under pressure starting at 1.4 GPa. As shown in Fig. 3(a), the sample crystallizes in the tetragonal T′ phase with space group *I*4/*mmm* for pressures ranging from 1.4 GPa to 20.7 GPa, which encompasses the critical pressure for the sign change in $R_{xy}(B)$. Given that the diffraction peaks of the T′ phase remain clearly visible and show systematic shifts to higher $2\theta$ values over the pressure range investigated, we propose that the sample remains in the T′ phase up to 20.7 GPa (Fig. 3(b)). We then extracted the lattice parameters $a$ and $c$ at each pressure point and plotted them in Fig. 3(c), revealing a continuous reduction with increasing pressure. These results confirm the absence of any pressure-induced structural phase transition within the studied pressure range. Thus, all changes observed in the high-pressure transport measurements are most likely of electronic origin.

It is worth noting that our room-temperature high-pressure XRD measurements reveal a different behavior from that observed at low temperature. Specifically, a partial transition from the T′ phase to the T phase starts at ~ 9.3 GPa (see SI). Upon further compression, the T′ phase and T phase coexist up to 41.4 GPa. The same phenomenon was also reported previously (34, 37). These observations suggest that the low-temperature T′ phase becomes metastable beyond the critical pressure ($P_{c1}$) as the temperature increases (see SI). At low temperatures, the system lacks sufficient thermal energy to overcome the energy barrier, thereby becoming trapped in the metastable T′ phase. However, as the temperature rises to room temperature, thermal fluctuations considerably affect its stability, driving the transformation toward the T phase (see SI).

**Pressure-Temperature Phase Diagram.** We summarize our experimental results for PLCCO in the pressure-temperature phase diagram (Fig. 4(a)), which is established from our resistance measurements on different samples (S#1-S#5, the data for S#5 is provided in the SI), along with the results from high-pressure Hall coefficient and XRD measurements at 40 K (Fig. 4). The diagram reveals that $T_c$ displays a small variation below $P_{c1}$ (~10 GPa), while the value of $R_H$ exhibits a continuous decrease from a large negative value (-12.7 $mm^3/C$) at ambient pressure to zero at $P_{c1}$ (Fig. 4(b)). At $R_H = 0$, the electron and hole carrier concentrations are balanced. Upon further compression, $R_H$ changes its sign from negative to positive. Since no crystal structure phase transition occurs at low temperature within the pressure range investigated (Fig. 4(c)), the high-pressure behavior observed in PLCCO is attributed to the pressure-induced reconstruction of the Fermi surface (FS). Such a Fermi surface reconstruction is commonly known as a Lifshitz transition (38, 39). The doping-induced Lifshitz transition has been observed in both bulk and film forms of hole-doped cuprates (40-42), and a similar doping-driven transition has also been reported in electron-doped cuprate films (43-45). However, no investigation of such a pressure-induced transition has yet been carried out. To the best of our knowledge, this is the first observation of a pressure-induced Lifshitz transition in the bulk electron- and hole-doped cuprate superconductors. Analysis of our high-pressure and low temperature XRD measurements find that the lattice parameter *c* of the sample is compressed by approximately 0.6 Å at $P_{c1}$, which is nearly six times larger than that of the lattice parameter *a* (~ 0.1 Å) at the same pressure. These results suggest that the Lifshitz transition observed here may be associated with the pressure-induced anisotropic lattice reductions.

Above $P_{c1}$, $T_c$ decreases monotonically with increasing pressure and eventually vanishes at $P_{c2}$ (taking the average of two adjacent points gives $P_{c2}$ = (17.0 GPa + 18.2 GPa)/2 = 17.6 GPa), beyond which the system enters a non-superconducting metallic ground state. This behavior is consistent with a superconducting (SC)-to-metallic (M) quantum phase transition (QPT). In the pressure range between $P_{c1}$ and $P_{c2}$, the Hall coefficient ($R_H$) remains positive and displays pressure-independent behavior (Fig. 4(b)). The observation that $R_H$ saturates while $T_c$ continues to decline suggests that the progressive suppression of superconductivity above $P_{c1}$ is not driven by further changes in carrier type or Fermi surface topology. Conversely, once the Lifshitz transition is complete and hole-type carriers dominate the system, further compression primarily diminishes the effective pairing interaction, resulting in a steady suppression of superconductivity. The phase diagram indicates that the Lifshitz transition at $P_{c1}$ and the SC-M QPT at $P_{c2}$ represent two pressure-driven electronic transitions in PLCCO.

**Theoretical Calculations and Discussion.** To better understand the correlation between superconducting transition temperature ($T_c$) and Hall coefficient ($R_H$) across the critical pressures in PLCCO, we employed an effective two-band model based on upper (UHB) and lower (LHB) Hubbard bands to calculate the evolution of band structure with pressure, incorporating the pressure-dependent parameters of the superconducting pairing strength ($J$) in the LHB and the band splitting ($R$) between the LHB and UHB (see SI). In this model, the superconducting order parameter is introduced only in the hole-pocket sector of the LHB, whereas the electron pocket does not participate directly in pairing. The FS reconstruction associated with $R$ compresses the hole-like Fermi-surface sector and thereby suppresses the superconducting gap. Given that pressure intrinsically enhances the orbital hopping integral $t$, we model $J/t$ and $R/t$ as continuously decreasing with pressure, while $R/t$ vanishes at $P_{c1}$, signaling a Lifshitz transition accompanied by FS reconstruction. The calculated zero-temperature superconducting gap $\Delta$ (in units of $|t_1|$) is shown in Fig. 5, with insets displaying the corresponding FS below and above $P_{c1}$. Below $P_{c1}$, electron- and hole-pockets coexist on the FS, where the essence of electron doping intrinsically enlarges the electron pocket, leading to a negative $R_H$. As pressure reduces $R$, the compression of the hole-like Fermi-surface sector is relieved, which tends to enhance superconductivity on the hole pocket. At the same time, the reduction of the pairing strength $J$ suppresses superconductivity. The competition between these two effects naturally stabilizes $T_c$ below $P_{c1}$. Beyond $P_{c1}$, the distinct electron- and hole- pockets fully merge into a single hole-dominated FS, resulting in a saturated positive $R_H$. Once this Lifshitz transition is completed, further pressure produces little additional change in Fermi-surface topology, whereas the continued reduction of $J$ continuously suppresses superconductivity until it vanishes entirely at $P_{c2}$. These trends are in excellent agreement with our experimental observations.

Although pressure drives superconductivity toward a QPT in both electron- and hole-doped cuprates, their non-superconducting ground states and QPT natures are markedly different. Hole-doped systems become insulating-like under pressure (29), while PLCCO in this study transitions to a metallic state. Additionally, the pressure-induced Lifshitz transition observed in electron-doped cuprates is absent in hole-doped ones (29). These differences extend to their fragility and decoherence: hole doping yields an insulating ground state with a resistance upturn at low temperatures, whereas electron doping results in a metallic state with saturated resistance. Together, these observations suggest that a full understanding of high-$T_c$ superconductivity requires considering not just pairing interactions, but also how superconductivity is destroyed and what follows. They also reveal that pressure does not simply act as a generic bandwidth-control parameter in cuprates. Its effect depends strongly on the initial electronic structure, carrier type, and lattice environment of the $CuO_2$ planes.

In conclusion, this work presents the first observation of a pressure-induced Lifshitz transition and superconducting-to-metallic quantum phase transition in the bulk electron-doped cuprate superconductor $Pr_{0.87}LaCe_{0.13}CuO_{4\pm\delta}$ (PLCCO) through high-pressure and low-temperature Hall and XRD measurements, along with high-pressure resistance measurements. Our results reveal a sign change in $R_H$ at $P_{c1}$ (~10 GPa), a clear signature of a Fermi surface reconstruction from electron- to hole-dominated topology, alongside only a slight variation in $T_c$. Low-temperature XRD measurements confirm the absence of any structural phase transition across $P_{c1}$, firmly establishing the purely electronic origin of this Lifshitz transition. Above $P_{c1}$, $T_c$ decreases and vanishes at $P_{c2}$ (~17.6 GPa), leading to a quantum phase transition from a superconducting state to a metallic state. Our findings demonstrate the pivotal role of Fermi surface topology in governing both the process of quantum phase transitions and the stability of superconductivity. Furthermore, they impose essential constraints for developing a unified theoretical description of the superconductivity in both the hole- and electron-doped cuprate families.

These authors with star (*) contributed equally to this work.

Correspondence and requests for materials should be addressed to: Shiliang Li (slli@iphy.ac.cn), Haozhe Liu (haozhe.liu@hpstar.ac.cn), Tao Xiang

(txiang@iphy.ac.cn), Richard L. Greene (rickg@umd.edu), and Liling Sun (liling.sun@hpstar.ac.cn or llsun@iphy.ac.cn).

## Acknowledgements

The work was supported by the National Key Research and Development Program of China (Grants No. 2021YFA1401800 and No. 2022YFA1403900) and the National Natural Science Foundation of China (Grants No. 12122414 and No. 12274207). The authors are grateful to the beam scientists at the Extreme Condition Methods of Analysis (EMA) beamline of the Brazilian Synchrotron Light Source (SIRIUS) and at the 15U beamline of the Shanghai Synchrotron Radiation Facilities for their assistance with the low and room temperature synchrotron XRD measurements conducted under high pressure. R.L.G. would like to acknowledge the support of the Maryland Quantum Materials Center. Partial work was supported by the Synergetic Extreme Condition User Facility (SECUF).

## Data Availability

The data that support the findings of this article are not publicly available. The data are available from the authors upon reasonable request.

## Competing interests

The authors declare no competing interests.

## Additional information

Supplementary information: The online version contains supplementary material available at https://xxx.

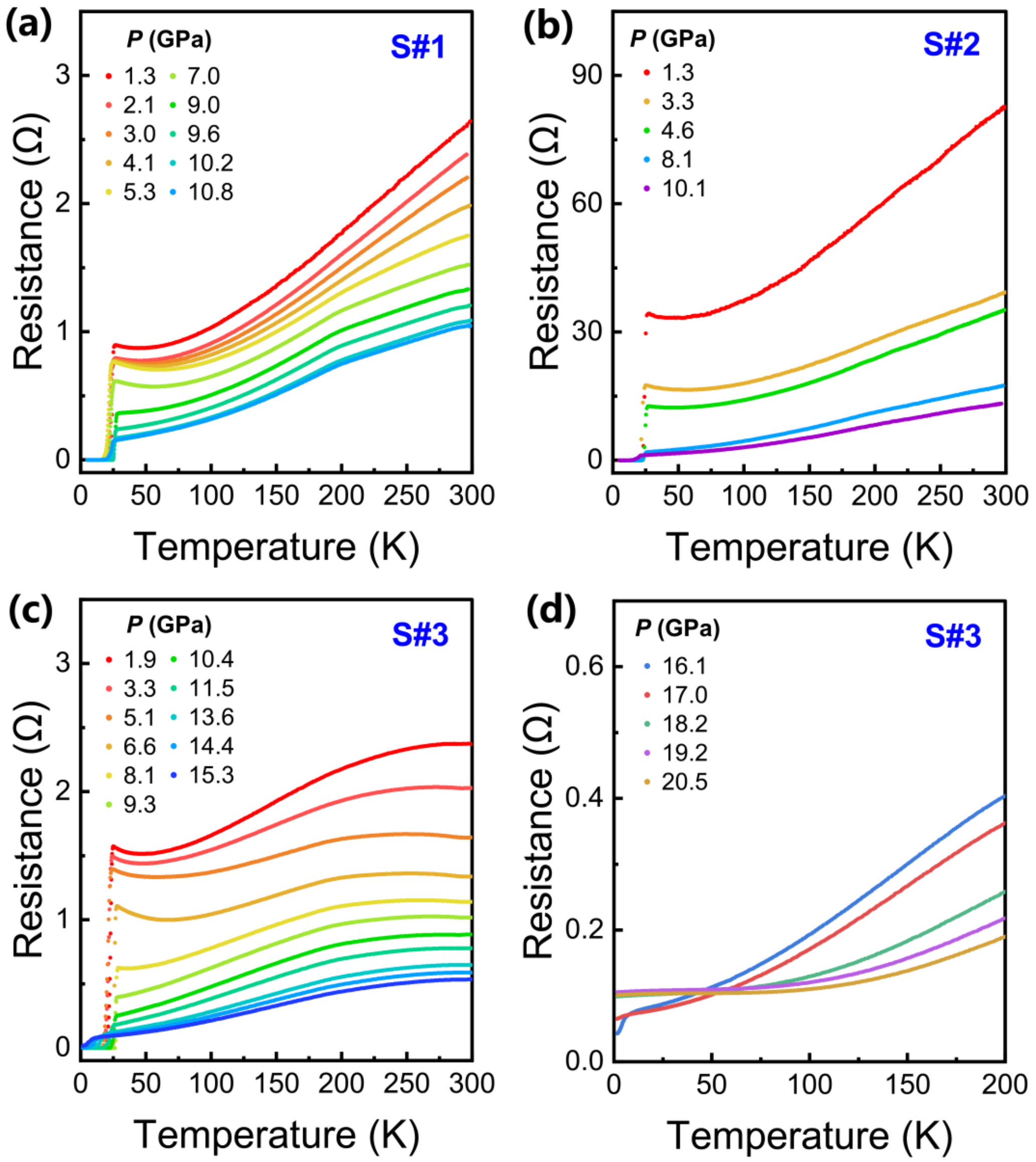

**Fig. 1. Transport properties of the $Pr_{0.87}LaCe_{0.13}CuO_{4\pm\delta}$ single crystals at different pressures.** Temperature dependence of *in-plane* resistance in the pressure range of (a) 1.3-10.8 GPa for sample 1 (S#1); (b) 1.3-10.1 GPa for sample 2 (S#2); (c) 1.9-15.3 GPa for sample 3 (S#3); (d) 16.1-20.5 GPa for sample 3 (S#3).

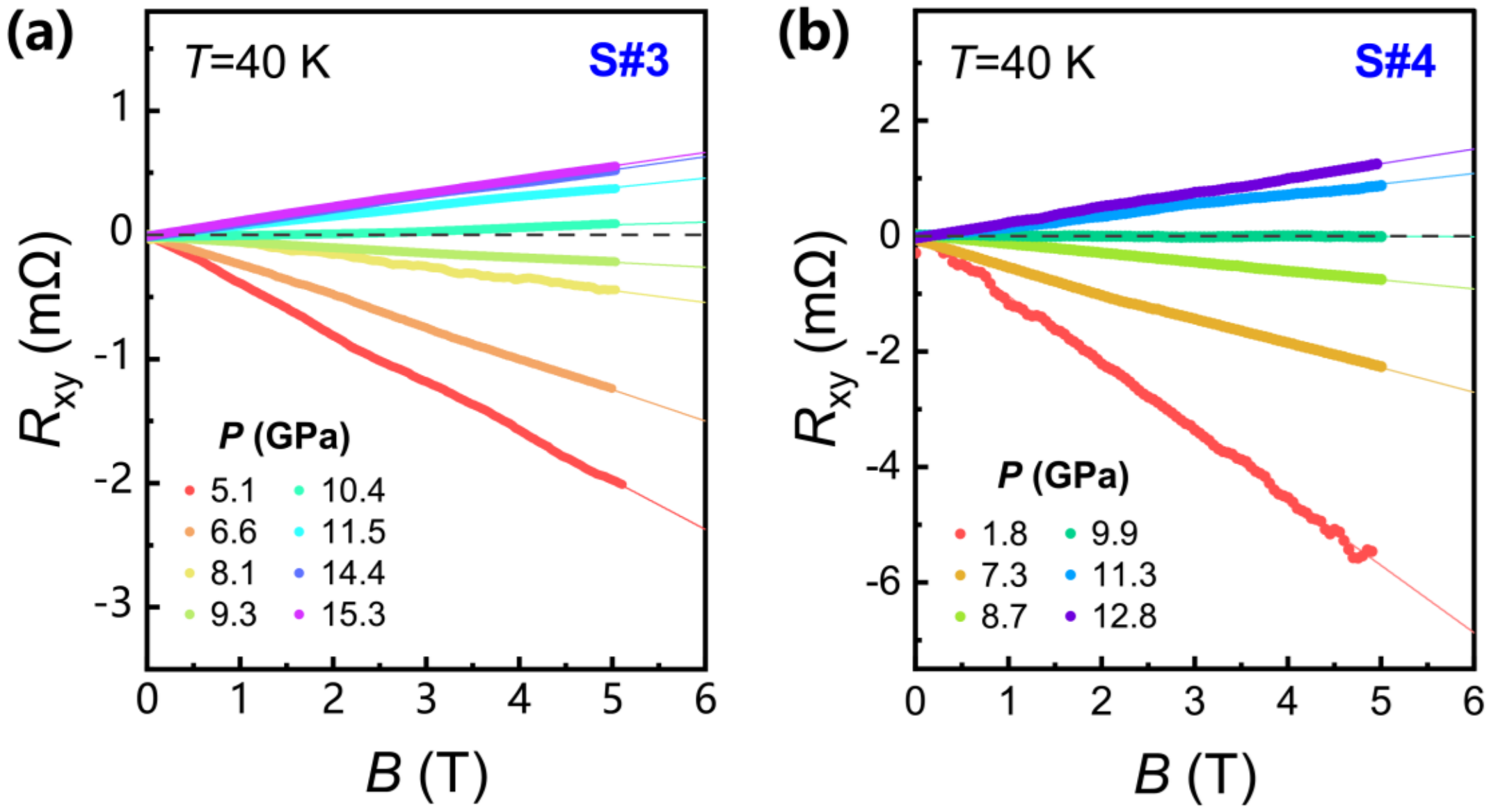

**Fig. 2. Hall resistance ($R_{xy}$) as a function of magnetic field ($B$) at 40 K for the $Pr_{0.87}LaCe_{0.13}CuO_{4\pm\delta}$ superconductors.** (a) $R_{xy}$ versus $B$ for sample 3 in the pressure range of 5.1-15.3 GPa. (b) $R_{xy}$ versus $B$ for sample 4 in the pressure ranges of 1.8-12.8 GPa. A crossover from an electron- to hole- dominant carrier behavior is observed in both samples.

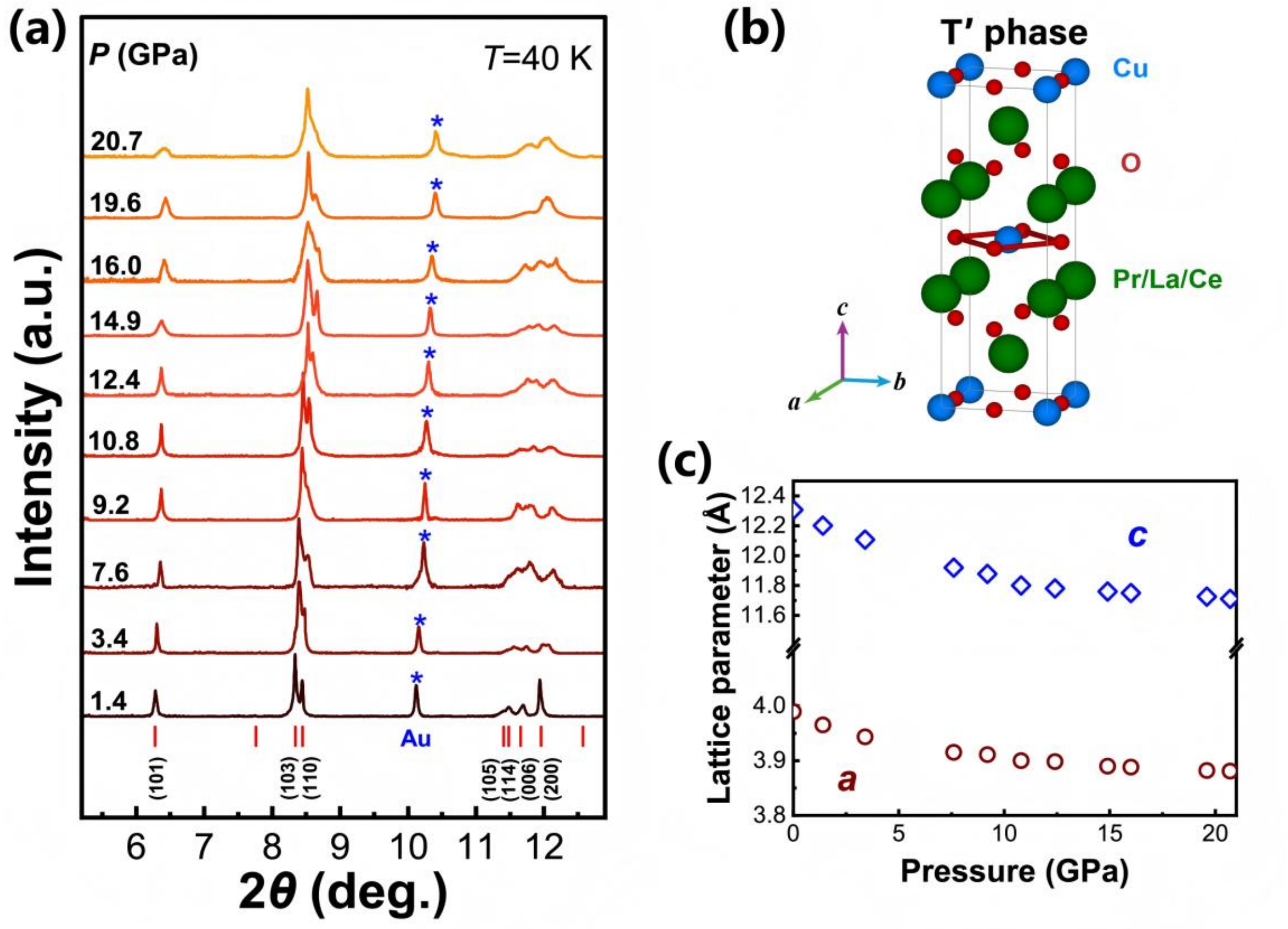

**Fig. 3. High-pressure structural information of $Pr_{0.87}LaCe_{0.13}CuO_{4\pm\delta}$ superconductors obtained at 40 K.** (a) X-ray diffraction patterns collected in the pressure range of 1.4-20.7 GPa. Peaks corresponding to the sample are indexed with Miller indices at the bottom. Given that gold (Au) was placed in the sample chamber to serve as an internal pressure calibrant, its diffraction peak (as indicated by the blue star) was observed. (b) Crystal structure of the tetragonal T′ phase, showing the positions of Cu (blue), O (red), and rare earth elements Pr/La/Ce (green) atoms in the tetragonal *I*4/*mmm* structure. (c) Pressure dependence of lattice parameters *a* and *c* for PLCCO in the T′ phase, displaying continuous compression of the lattice parameters with increasing pressure up to 20.7 GPa. The ambient-pressure data in Fig. 3(c) is extrapolated from our high-pressure and low-temperature measurements.

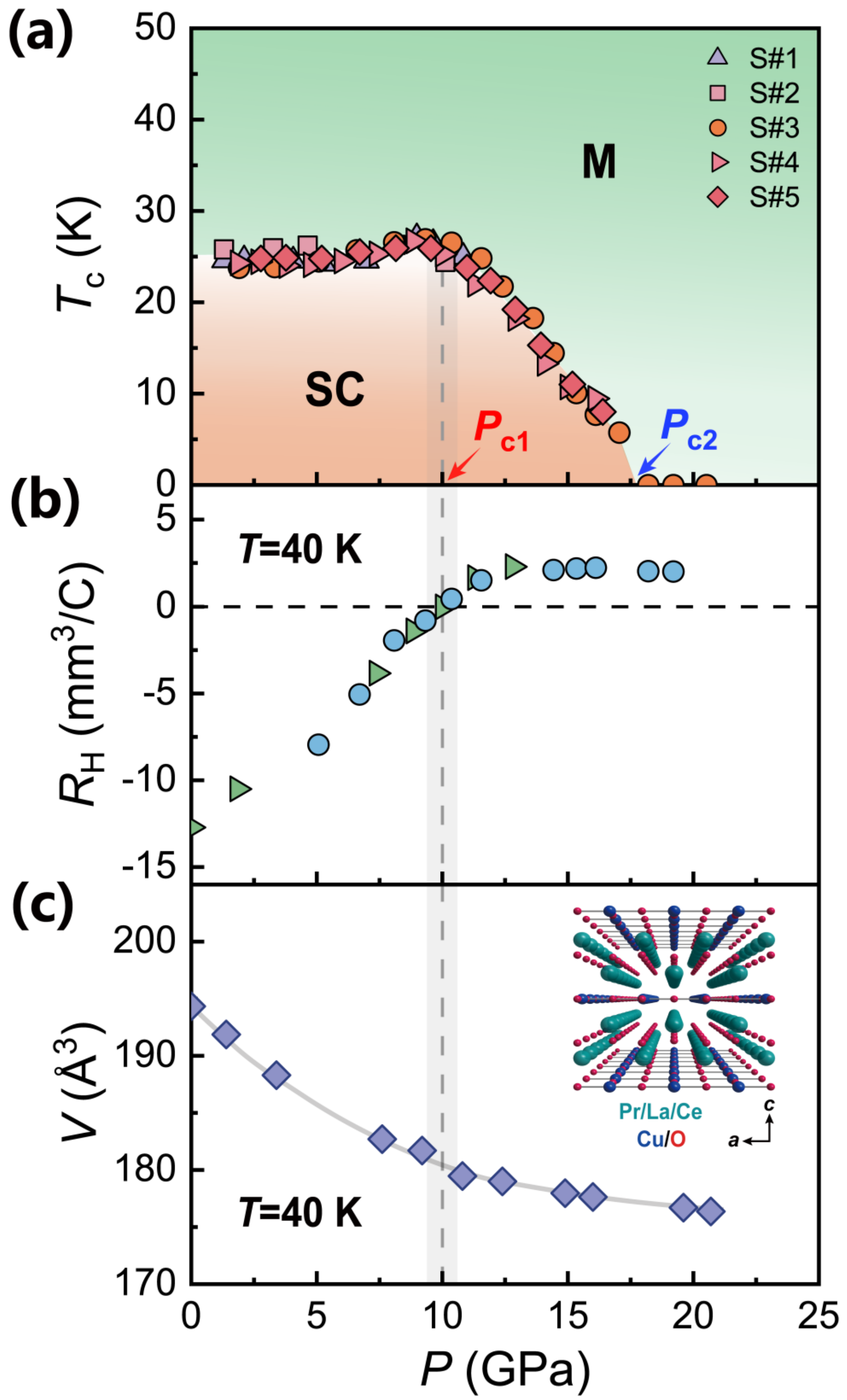

**Fig. 4. Pressure dependence of superconducting transition temperature ($T_c$), Hall coefficient ($R_H$), and unit cell volume ($V$) for $Pr_{0.87}LaCe_{0.13}CuO_{4\pm\delta}$ superconductors.**

(a) $T_c$ as a function of pressure for five PLCCO samples (S#1-S#5). Below $P_{c1}$ (~10 GPa), $T_c$ exhibits a small variation with increasing pressure. Above $P_{c1}$, a clear suppression of superconductivity is observed, with $T_c$ decreasing monotonically from ~ 25 ± 1.5 K at 10 GPa down to 10 K at 15.3 GPa. At $P_{c2}$ (~17.6 GPa), $T_c$ disappears completely and a metallic state sets in. (b) Pressure dependence of Hall coefficients ($R_H$) obtained at 40 K, showing a sign change from a large negative value (-12.7 mm$^3$/C, estimated by an extrapolation of the data from S#4) at ambient pressure to a positive value above $P_{c1}$, and then shows a steady state between $P_{c1}$ and $P_{c2}$. (c) Unit cell volume ($V$) as function of pressure at 40 K, showing continuous compression, from ~ 195.9 Å$^3$ at ambient pressure to ~ 176.4 Å$^3$ at 20.7 GPa, without structural phase transitions. The inset of Fig.4(c) illustrates the schematic crystal structure of PLCCO in the T′ phase. The vertical dashed line marks the critical pressures ($P_{c1}$) where a Lifshitz transition takes place. The ambient-pressure data in Figs. 4(b) and 4(c) are extrapolated from our high-pressure and low-temperature measurements.

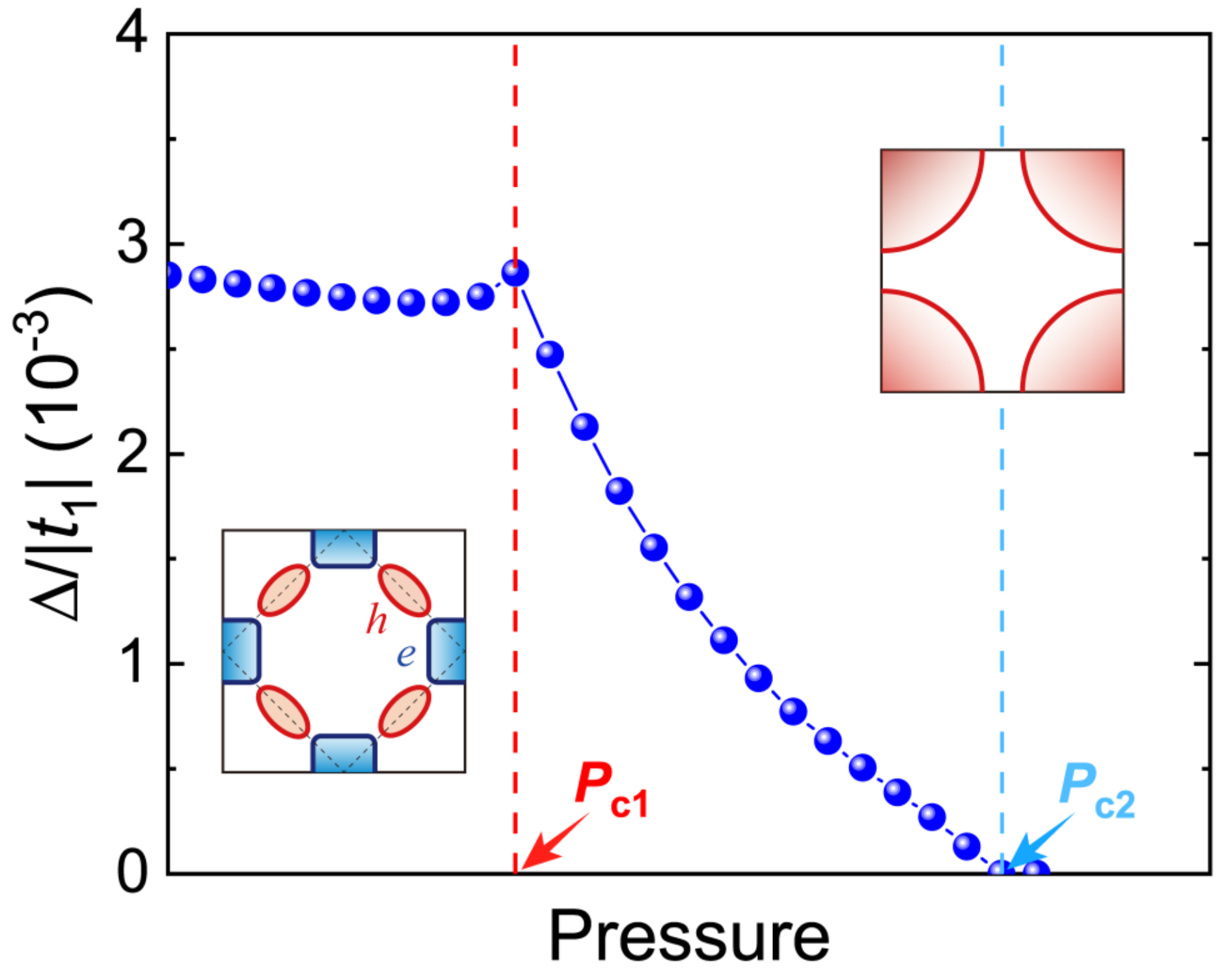

**Fig. 5. Results of theoretical calculations.** The variation of zero-temperature superconducting gap $\Delta$ (in units of $|t_1|$) across a Lifshitz transition at $P_{c1}$ and quantum phase transition at $P_{c2}$. The insets display the corresponding electron- (blue) and hole- (red) type pockets on FS below and above $P_{c1}$.